\documentclass[aps,prl,twocolumn,showpacs,groupedaddress,floatfix]{revtex4}
\usepackage{graphicx,color,graphics,caption,epsf,dcolumn}
\usepackage{times,nicefrac,amsmath,amsfonts,amssymb,amsthm,bm,bbm}

\begin{document}
\title{Condensational theory of stationary tornadoes}

\author{A.~M.~Makarieva, V.~G.~Gorshkov, and A.~V.~Nefiodov}
\email[Corresponding author: ]{anef@thd.pnpi.spb.ru}
\affiliation{Theoretical Physics Division, Petersburg Nuclear
Physics Institute, 188300 Gatchina, St.~Petersburg, Russia}

\begin{abstract}

Using the Bernoulli integral for air streamline with condensing
water vapor a stationary axisymmetric tornado circulation is
described. The obtained profiles of vertical, radial and tangential
velocities are in agreement with observations for the Mulhall
tornado, world's largest on record and longest-lived among the three
tornadoes for which 3D velocity data are available. Maximum possible
vortex velocities are estimated.
\end{abstract}

\pacs{47.10.-g, 47.90.+a, 47.45.-n, 51.90.+r}

\maketitle

\section{The condensational pressure potential}

Tornado circulation induced by water vapor condensation can be
described as follows. Condensation of water vapor in the
adiabatically ascending air results in a drop of air pressure by
$\Delta p = p_v$, where $p_v$ is water vapor pressure at the Earth's
surface. The decrease of pressure along the vertical axis sustains
the ascending air motion with vertical velocity $w$ and induces a
compensating horizontal air inflow with radial velocity $u$. The
converging radial flow has maximal velocity at the surface, where
the magnitude of the condensation-induced pressure drop is the
largest. Radial velocity approaches zero at a certain height $z =
h$, which approximately coincides with the cloud height. In the
upper atmosphere at $z > h$ the condensed water is transported away
from the condensation area by the strong updraft and outgoing air
flow. It precipitates at a considerable distance from the center of
the condensation area.

The continuity equation in the cylindrical system of coordinates
relates radial $u$ and vertical $w$ velocities of the axially
symmetrical vortex as $w = (h/r)(\partial ur/\partial r)$. The
vertical and horizontal pressure gradient forces induced by
condensation are $\Delta p/h$ and $\partial p/\partial r$,
respectively. Equating the power of the vertical and radial air
flow, $u \,\partial p/\partial r = w(\Delta p/h)$, and accounting
for the continuity equation, we obtain $\partial p/\partial r =
\Delta p(ur)^{-1} (\partial ur/\partial r)$. This corresponds to
pressure potential $p = \Delta p \ln ur + \mathrm{const}$ \cite{mg09b}.
Its exact derivation is given in work \cite{mg11}.

\section{The Bernoulli integral for condensation-induced tornadoes}

For the high wind velocities of intense vortices to arise, the
condensational pressure gradients within both tornadoes and
hurricanes must significantly exceed turbulent friction. In such a
case, the Euler equations possess a Bernoulli integral for the
streamline:
\begin{gather}
B(r) \equiv \frac{1}{2}\rho(u^2 + w^2 + v^2) + \Delta p \ln ur =
B(r_1)\,  , \label{eq1}\\
p(r) = p_1 + \Delta p \ln \frac{ur}{u_1 r_1} \, , \quad \Delta p = p_v \equiv \rho
\frac{u_c^2}{2} \, , \label{eq2}\\
p_1 \equiv p(r_1) \,  , \quad  w = \frac{h}{r} \frac{\partial
ur}{\partial r}\,  ,  \quad v = \frac{a}{r}\, .    \label{eq3}
\end{gather}
Here $u$, $w$ and $v$ are the radial, vertical and tangential
velocities, respectively, $r$ is distance from the center of the
condensation area, $r = r_1$ is the outer border of the condensation
area, $u_1 \equiv u(r_1)$, $\rho$ is air density, $u_c$ is the velocity scale determined
by water vapor condensation, $a$ is angular momentum per unit air
mass, and $z < h$ is the region of converging streamlines ($u > 0$).

It is convenient to use the following units
\begin{equation}\label{eq4}
\Delta p = 1\, , \quad u_c = 1\, , \quad \rho = 2\, , \quad r_1 = 1.
\end{equation}
In these units, the Bernoulli integral and the pressure potential
become dimensionless
\begin{eqnarray}
B(r)-B(r_1)&=& u^2 - u^2_1 + w^2 - w^2_1 \nonumber\\
&&  +\left(\frac{a^2}{r^2} - a^2 +
\ln \frac{u r}{u_1} \right)=0 \,  , \label{eq5}\\
p(r) &=& p_1 +   \ln \frac{ur}{u_1} \, , \label{eq6}
\end{eqnarray}
where  $w_1 \equiv w(r_1)$ and $a \equiv  v_1 \equiv  v(r_1)$.

Let us introduce a new variable $y \equiv  ur/u_1$. Then
Eq.~\eqref{eq5} takes the form of a nonlinear differential equation
on $y$:
\begin{gather}
y' = \frac{r}{u_1 h} \sqrt{u_1^2\left(h^2{y_1'}^2 + 1 -
\frac{y^2}{r^2}\right) - \left(\frac{a^2}{r^2}-a^2 +\ln y\right)}, \label{y}\\
w \equiv u_1\frac{h}{r}y'\, , \quad u \equiv u_1 \frac{y}{r}\, ,
\quad v = \frac{a}{r}\,  ,\quad p=p_1+\ln y \,  , \\
y\equiv\frac{ur}{u_1}\, ,\quad y'\equiv \frac{dy}{dr}\, , \quad
y_1'\equiv y'(r_1) \, .
\end{gather}
Real solution of Eq.~\eqref{y} exists at those $r$ only, where the
expression under the square root in Eq.~\eqref{y} is positive. The
internal radius $r = r_0$, where condensation ceases, is obtained by
equating the last term in the round brackets in Eqs.~\eqref{eq5}
and~\eqref{y} to zero at $y(r_0) = r_0$, which is equivalent to
$u(r_0) = u_1$. Condensation commences at $r_1$ and ceases at $r_0$
at one and the same radial velocity $u_1$. As follows from
Eq.~\eqref{y}, the following relationships are simultaneously
satisfied:
\begin{gather}
\frac{a^2}{r_0^2} - a^2 + \ln r_0 = 0 \, ,
\quad  y_0\equiv y(r_0)=r_0\, , \label{r0}\\
y_1 \equiv y(r_1) =1 \,  , \quad y_0' \equiv  y'(r_0)=r_0y_1'\, , \label{y0}\\
u_0 \equiv u(r_0) = u_1\, , \quad w_0 \equiv w(r_0) = w_1= u_1 h
y_1' \,  .
\end{gather}
At $r = r_0$  all condensational
potential energy is converted to the kinetic energy of rotation.
At $r < r_0$, real solutions of Eq.~\eqref{y} for velocities $u$,
$v$, and $w$ do not exist.

Equation~\eqref{y} is a first-order differential equation with one
boundary condition: $y_1 = 1$ ($u(r_1) = u_1$) or $y_0 = r_0$
($u(r_0)=u_1$).  If at fixed $y_1$ one considers the constant $y_1'$
in Eq.~\eqref{y} as a free parameter, then in the general case the
interval, where real solutions exist, does not include the point $r
= r_0$, which means that the maximum velocity $v_\mathrm{max} \sim
a/r_0$ is not reached and the tornado does not exist.

Tornado exists, when the interval of real solutions comprises the
point $r_0$ defined by Eq.~\eqref{r0}. Solution of Eq.~\eqref{y},
that is real within the range $r_0 \leqslant r \leqslant 1$,  is
obtained by setting the boundary condition on $y$ at $r_0$ as $y_0 =
r_0$ and choosing $y_1'$ at given $u_1$, $a$ and $h$ such that $y_1 = 1$.

\section{Comparison with observations}

Data of three-dimensional circulation  (the dependencies of the
velocities $u$, $w$ and $v$ on distance $r$ from the tornado center)
have only recently become available and exist for three tornadoes
\cite{kos10,kos08,lee05}. We shall consider the Mulhall tornado
(Oklahoma, 3 May 1999), which is the longest-lived (1 hour 20 min
\cite{spe02}) and longest-observed (18 min \cite{lee05}) among the
three as well as world's largest tornado on record \cite{lee05}.

According to empirical observations, the intense tornadoes can
occur, when the mean relative humidity at $z \lesssim 1$~km is not
lower than 75-85\% \cite{tho03}. At a cha\-rac\-te\-ris\-tic surface
temperature 30~$^\mathrm{o}$C \cite{tana06} and 80\% relative
humidity the vapor pressure at the surface is $p_v = \Delta p \simeq
30$~hPa. Taking air density $\rho = 1.15$~kg~m$^{-3}$ in
Eq.~\eqref{eq2}, we obtain the characteristic velocity $u_c =
73$~m~s$^{-1}$. Velocities $v_1$ and $u_1$ at the external border $r
=r_1$ must be the functions of translational velocity $U$ (speed of
movement of tornado as a whole). We put radial velocity $u_1 =
U/\pi$ \cite{mg11}, taking into account that the flux of moist air
via tornado cross-section $2r_1U$ is equal to the flux via tornado
circumference $2\pi r_1 u_1$. We put tangential velocity $v_1 =
2U/\pi$ assuming that the angular momentum of the main streamline that
delivers moist air into the condensation area (see, e.g., Fig.~8 in
work \cite{wur10}) is determined by the mean value of $U \cos
\alpha$. Here $0 \leqslant \alpha  \leqslant \pi/2$ is a random
angle between velocity at this streamline and radius-vector $r$ at
the point $r = r_1$, where the air enters the condensation area.
From $U = 13$~m~s$^{-1}$ \cite{lee05} we have for dimensionless
variables $u_1 = U/\pi u_c = 0.06$, $v_1 = 2U /\pi u_c = 0.12$. For
$a = v_1=0.12$, we obtain the eye radius $r_0 = 0.074$ from
Eq.~\eqref{r0}. Taking cloud height $h=1.2$~km and total size of
tornado condensation area $r_1 = 8.5$~km, we have dimensionless
value $h = 0.14$.

\begin{figure}[tbh]
\centerline{
\resizebox{\columnwidth}{!}{%
\includegraphics{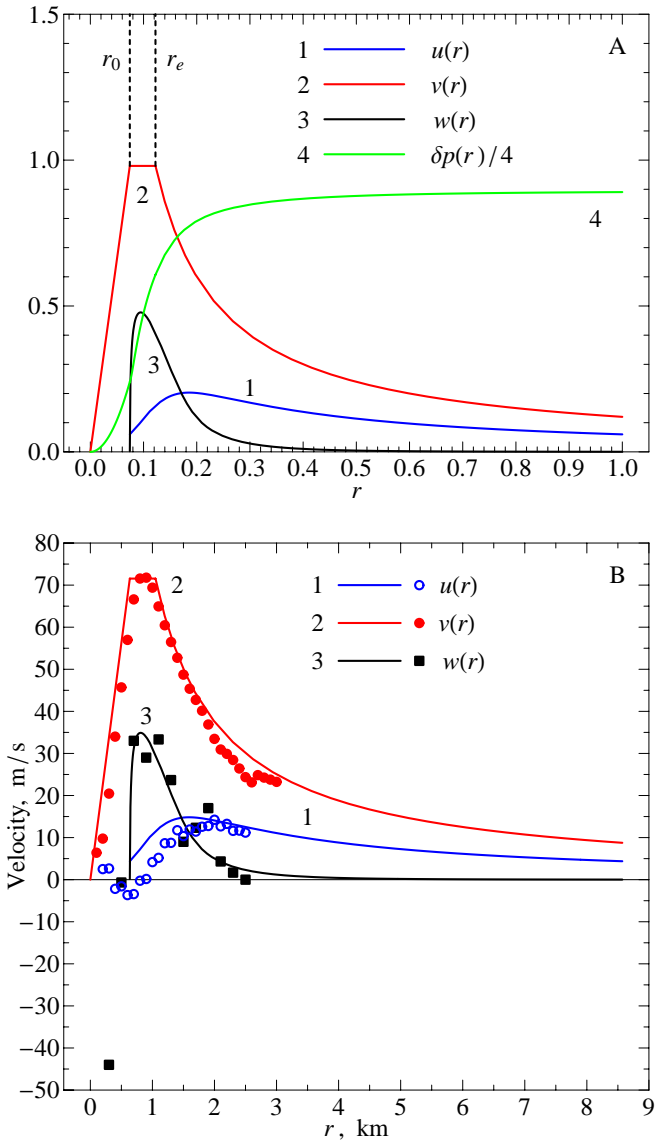}}}
\caption{\label{fig1}(color online). A: Solution of Eq.~\eqref{y} at
$a=0.12$ ($r_0 = 0.074$), $u_1 = 0.06$, $h = 0.14$ in dimensionless
units \eqref{eq4}, $\delta p(r) \equiv p(r) - p(0)$. B: Comparison
with observations for the Mulhall tornado \cite{lee05} at $u_c =
73$~m~s$^{-1}$, $h = 1.2$~km ($r_1 = 8.5$~km). The negative vertical
velocity (downdraft) within the tornado eye and the decrease of
radial velocity near $r_0$ is related to non-stationarity of eye
rotation not described by the Bernoulli integral \eqref{y}, the
latter pertaining to the converging ascending streamline. }
\end{figure}

For these particular parameters the numerical solution of
Eq.~\eqref{y} obtained by using conditions \eqref{r0} and \eqref{y0}
corresponds to $y_1' = 0.03574$ (see Fig.~\ref{fig1}A). The account
of stationary eye rotation is made in the same way as in work
\cite{mg11}, when a certain part of tangential kinetic energy
developed in the condensation area is spent on solid-body rotation
and creation of the pressure gradient in the eye of radius $r_0$.
This lowers tangential velocity in the transitional region $r_0
\leqslant r < r_e$ between the condensation area and the eye, where
$r_e = 1.65 r_0$ \cite{mg11}. The expressions for tangential
velocity and pressure at $r < r_0$ coincide with Eqs.~(23)--(25) in
work \cite{mg11}. The em\-pi\-ri\-cal points shown in
Fig.~\ref{fig1}B characterize the Mulhall tornado close to the time
of peak intensity. They correspond to cha\-rac\-te\-ris\-tic
vertical velocity $w(r)$ at $z = 650$~m \cite[Fig.~4b]{lee05}, mean
radial velocity $u(r)$ at $150\mbox{ m} < z < 850$~m
\cite[Fig.~4b]{lee05} and mean tangential velocity $v(r)$ at
$50\mbox{ m} < z < 950$~m \cite[Fig.~5a]{lee05}.

\noindent

It is seen from Fig.~\ref{fig1}B that to the right side of the
maximum the radial distribution of mean tangential velocity at $z
\leqslant h$ conforms well to the assumption of conserved angular
momentum \eqref{eq3}. The choice of $h =1.2$~km is supported by
observation that in this layer the radial velocity $u(r)$ exceeds
$u_1 = 4.4$~m~s$^{-1}$ over considerable part of tornado circulation
\cite[Fig.~4b]{lee05}. (It should be noted that two other
tornadoes, for which the data are available \cite{kos10,kos08}, have
a significantly lower inflow level, $h \leqslant 400$~m, than the
Mulhall tornado. Mean tangential velocities do not follow the
conserved angular momentum distribution. The decrease of angular
momentum towards the center demands a more detailed consideration
with additional parameters \cite{mg11}).

Total pressure fall as shown in Fig.~\ref{fig1}A is $3.6 \Delta p =
108$~hPa. This is in agreement with the few available direct
measurements of tornado surface pressure. In the Manchester tornado
(South Dacota, 2003), which was of the same (F4) intensity as the
Mulhall tornado, a pressure fall of 100~hPa was registered
\cite{lee05}.

\section{Conditions of vortex existence and the maximum possible velocities}

\begin{figure}[thp]
\centerline{
\resizebox{\columnwidth}{!}{%
\includegraphics{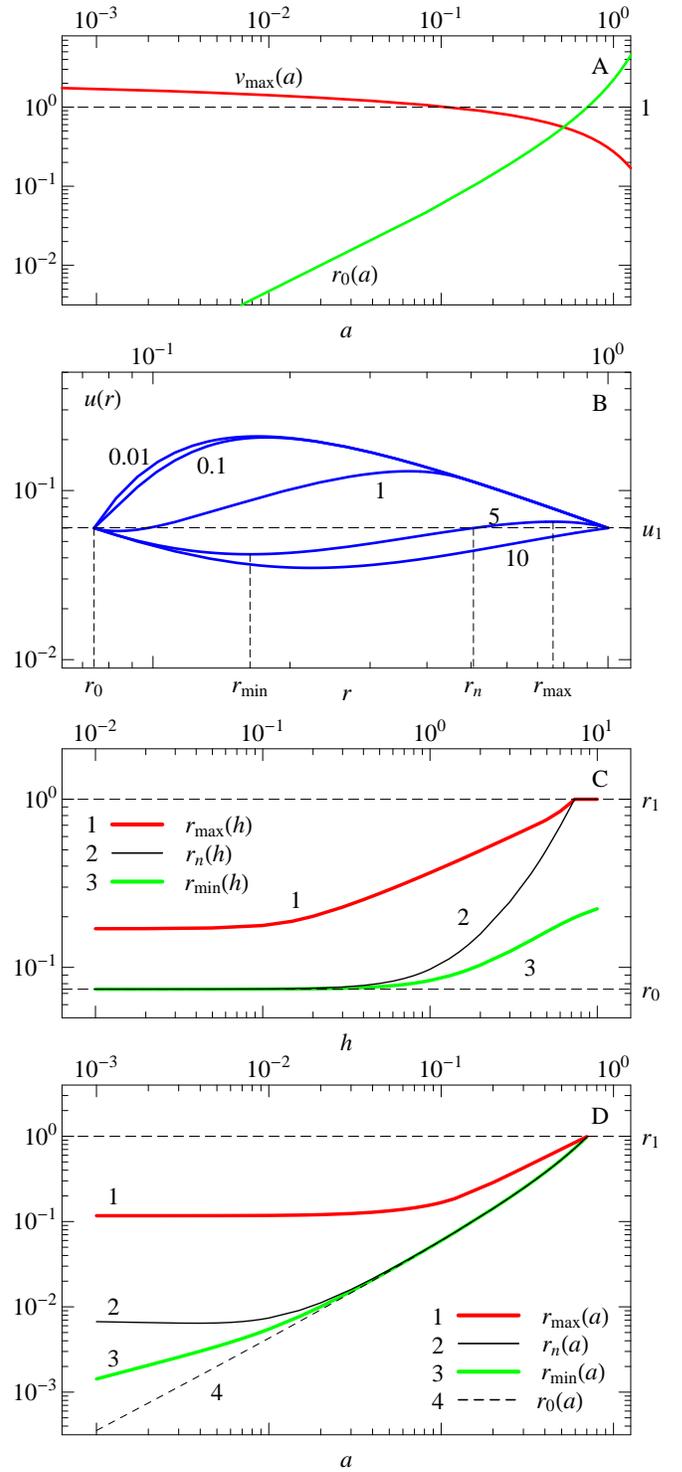}}}
\caption{ \label{fig2}(color online). A: Dependence of eye radius
$r_0$ and maximal tangential velocity $v_\mathrm{max} = a/r_e$, $r_e
= 1.65 r_0$ \cite{mg11} on angular momentum $a$ in dimensionless
units \eqref{eq4}. B: Radial velocity $u(r)$ obtained by solving
Eq.~\eqref{y} under conditions \eqref{r0} and \eqref{y0} at $a =
0.12$, $u_1 = 0.06$ for five values of $h$ (shown near the curves).
Points $r_{\mathrm{min}}$, $r_{\mathrm{max}}$  and $r_n$  shown for
$h = 5$ correspond to radial velocity minimum, maximum and $u(r_n) =
u_1$, respectively. C: Dependence of $r_{\mathrm{min}}$,
$r_{\mathrm{max}}$ and $r_n$ on $h$ at $a=0.12$, $u_1 = 0.06$. D:
Dependence of $r_{\mathrm{min}}$, $r_{\mathrm{max}}$ and $r_n$ on
$a$ at $h=0.14$, $u_1 = 0.06$; $r_0(a)$ (\ref{r0}).}
\end{figure}

With account of stationary eye rotation \cite{mg11} the maximum wind
velocity $v_\mathrm{max} = a/r_e$ (and, correspondingly, the maximum
kinetic energy) is achieved at $r_e = 1.65 r_0$ (see
Fig.~\ref{fig1}A), where $r_0$ is a function of $a$ given by
Eq.~\eqref{r0}, Fig.~2A. The Earth rotation does not determine
angular momentum in tornado due to the small linear size of the
vortex. The value of $a$ is related to the translational velocity
$U$. This velocity cannot be infinitely small: tornado exists at the
expense of water vapor accumulated in the atmosphere and, hence,
must move to sustain itself \cite[pp.~227-229]{g95}. Maximum
velocity attainable in the condensational vortex depends only weakly
on angular momentum and grows rather slowly (logarithmically) with decreasing $a$
(see Fig.~\ref{fig2}A). For realistic $a
\geqslant10^{-3}$ ($v_1 \geqslant 0.1$~m~s$^{-1}$), $v_\mathrm{max}$
does not exceed $1.7u_c \simeq 120$~m~s$^{-1}$. This agrees well
with the available estimates of maximum wind speeds in tornadoes
\cite{lee05}.

\noindent

The existence of vortex is related to a certain minimum value of
radial velocity $u$, which describes the atmospheric inertia with
respect to the development of condensational circulation. At $u <
u_1$ condensation ceases. The condition $u_0 = u_1$ corresponds to
the following relationships
\begin{eqnarray}
u'(r_0) &=& \frac{u_0}{r_0} (y_0' - 1) = u_1 \left(y_1' -
\frac{1}{r_0}\right) \simeq -\frac{u_1}{r_0}\,  ,  \label{u0}\\
u'(r_1) &=& u_0 \left(\frac{y_0'}{r_0}  - 1\right) = u_1 \left(y_1'
- 1\right) \,  .
\end{eqnarray}
This means that there is a minimum of $u(r)$ at $r=r_{\mathrm{
min}}$ within  $r_0 < r_{\mathrm{ min}} < r_1$. The existence of
condensation at $u(r) > u_1$ means that there is also a point
$r=r_\mathrm{max}$, $r_0 < r_\mathrm{max} < r_1$, where $u(r)$ is
maximum. It follows that there is a point $r = r_n$, $r_0 <
r_\mathrm{min} < r_n < r_\mathrm{max} < r_1$, where $u(r_n) = u_1$.
At $r < r_n$ there is no condensation and no condensational pressure
potential to accelerate air. When $r_n \gg r_0$ the maximum vortex
velocity, $a/r_e$, is not reached: $v_\mathrm{max} = a/r_n \ll
a/r_e$. Therefore, tornado exists, if the following condition
$\kappa \equiv (r_n - r_0)/r_0 \ll 1$ is fulfilled ($\kappa \simeq
10^{-3}$ for the vortex shown in Fig.~\ref{fig1}).

Analysis of Eq.~\eqref{y} shows that this condition is violated with
increasing $h$, which, at a fixed height of the atmosphere,
corresponds to diminishing linear size $r_1$ of the condensation
area. In Fig.~\ref{fig2}B, profiles of $u(r)$ are shown for $h$
varying from $0.01$ to $10$. At $h \geqslant 7.3$ we have $y'_1
\geqslant1$, $u'(r_1) \geqslant 0$ and maximum of $u(r)$ at $r <
r_1$ disappears. Decreasing $a$ at fixed $h$ also leads to
increasing $\kappa$, Fig.~2D. It follows that the smaller the
horizontal size of the condensation area, the larger the angular
momentum that is needed for a vortex to arise. A given value of
angular momentum sets the minimal horizontal size of the vortex. For
the parameters shown in Fig.~\ref{fig1} the minimum possible vortex,
where velocity $v \simeq u_c \simeq 70$~m~s$^{-1}$ can be observed,
corresponds to $h \sim 1$ (see Fig.~\ref{fig2}C). The minimal
condensation area has then radius $r_1 \sim 1.2$~km and funnel (eye)
radius of about 90~m. At small $a$ and $r_1$ only ordinary squalls
can develop.


\begin{thebibliography}{99}
\bibitem[1]{mg09b}
A.~M.~Makarieva and V.~G.~Gorshkov, \newblock Phys. Lett. A {\bf
373}, 4201 (2009).
\bibitem[2]{mg11}
A.~M.~Makarieva and V.~G.~Gorshkov,  \newblock  Phys. Lett. A {\bf
375}, 1053 (2011).
\bibitem[3]{kos10}
K.~Kosiba and  J.~Wurman,  \newblock  J. Atmos. Sci. {\bf 67}, 3074
(2010).
\bibitem[4]{kos08}
K.~A.~Kosiba, R.~J.~Trapp, and J.~Wurman, \newblock Geophys. Res.
Lett. {\bf 35}, L05805 (2008).
\bibitem[5]{lee05}
W.-C.~Lee and J.~Wurman, \newblock J. Atmos. Sci. {\bf 62}, 2373
(2005).
\bibitem[6]{spe02}
D.~A.~Speheger,  C.~A.~Doswell, and G.~J. Stumpf, \newblock Wea.
Forecast. {\bf 17}, 362 (2002).
\bibitem[7]{tho03}
R.~L.~Thompson,  R. Edwards, J.~A. Hart, K.~L. Elmore, and P.
Markowski, \newblock  Wea. Forecast.  {\bf 18}, 1243 (2003).
\bibitem[8]{tana06}
R.~L.~Tanamachi,  H.~B. Bluestein, S.~S. Moore, and R.~P. Madding,
\newblock  J. Atmos. Oceanic Technol. {\bf 23}, 1445 (2006).
\bibitem[9]{wur10}
J.~Wurman, K. Kosiba, P. Markowski, Y. Richardson, D. Dowell, and P.
Robinson, \newblock Mon. Wea. Rev. {\bf 138}, 4439 (2010).
\bibitem[10]{g95}
V.~G.~Gorshkov, \newblock {\em Physical and Biological Bases of Life
Stability} \newblock (Springer, Berlin, 1995).
\end{thebibliography}
\end{document}